\documentclass[%
 aip,cha,
 amsmath,amssymb,
 reprint,%
]{revtex4-1}

\usepackage{graphicx}
\usepackage{dcolumn}
\usepackage{bm}

\usepackage[utf8]{inputenc}
\usepackage[T1]{fontenc}
\usepackage{mathptmx}
\usepackage{etoolbox}
\usepackage{orcidlink}

\newcommand{\Qdim}{_\text{Q}^\text{dim}}  
\newcommand{\Qreg}{_\text{Q}^\text{reg}} 
\newcommand{\Qfull}{_\text{Q}^\text{full}} 
\newcommand{\FN}{_\text{FN}} 
\newcommand{\FNdim}{_\text{FN}^\text{dim}} 
\newcommand{\FNreg}{_\text{FN}^\text{reg}} 
\newcommand{\FNfull}{_\text{FN}^\text{full}} 

\makeatletter
\def\@email#1#2{%
 \endgroup
 \patchcmd{\titleblock@produce}
  {\frontmatter@RRAPformat}
  {\frontmatter@RRAPformat{\produce@RRAP{*#1\href{mailto:#2}{#2}}}\frontmatter@RRAPformat}
  {}{}
}%
\makeatother
\begin{document}

\preprint{AIP/123-QED}

 \title[]{Dynamical equivalence between resonant translocation of a polymer chain and diversity-induced resonance}

 \author{Marco Patriarca\,\orcidlink{0000-0001-6743-2914}}
\email{marco.patriarca@gmail.com}
 \affiliation{National Institute of Chemical Physics and Biophysics, Akadeemia Tee 23, 12618 Tallinn, Estonia \looseness=-1}
 \affiliation{Department of Cybernetics, Tallinn University of Technology, Ehitajate tee 5, 19086 Tallinn, Estonia \looseness=-1}

 \author{Stefano Scialla\,\orcidlink{0000-0003-1582-8743}}
   \affiliation{National Institute of Chemical Physics and Biophysics, Akadeemia Tee 23, 12618 Tallinn, Estonia \looseness=-1}
   \affiliation{Department of Engineering, Università Campus Bio-Medico di Roma, Via Á. del Portillo 21, 00128 Rome, Italy \looseness=-1}

 \author{Els Heinsalu}
 \affiliation{National Institute of Chemical Physics and Biophysics, Akadeemia Tee 23, 12618 Tallinn, Estonia \looseness=-1}

 \author{Marius E. Yamakou\,\orcidlink{0000-0002-2809-1739}}
 \affiliation{Department of Data Science, Friedrich-Alexander-Universit\"{a}t Erlangen-N\"{u}rnberg, Cauerstr. 11, 91058 Erlangen, Germany \looseness=-1}

 \author{Julyan H.E. Cartwright\orcidlink{0000-0001-7392-0957}}
 \affiliation{Instituto Andaluz de Ciencias de la Tierra, CSIC, 18100 Armilla, Spain \looseness=-1
 }
 \affiliation{Instituto Carlos I de Física Teórica y Computacional, Universidad de Granada, 18071 Granada, Spain \looseness=-1}

\date{\today}
             
\begin{abstract}
Networks of heterogeneous oscillators are often seen to display collective synchronized oscillations, even when single elements of the network do not oscillate in isolation.
It has been found that it is the diversity of the individual elements that drives the phenomenon, possibly leading to the appearance of a resonance in the response.
Here we study the way in which heterogeneity acts in producing an oscillatory regime in a network and show that the resonance response is based on the same physics underlying the resonant translocation regime observed in models of polymer diffusion on a substrate potential. 
Such a mechanical analog provides an alternative viewpoint that is useful to interpret and understand the nature of collective oscillations in heterogeneous networks.
\end{abstract}

\maketitle

\begin{quotation}
Many networks of cells in the human body, including neurons~\cite{Soriano-2023a}, $\beta$-cells in the pancreatic islets of Langerhans~\cite{Peercy-2022a}, and the cardiomyocytes of the heart muscle~\cite{Maltsev-2022a}, present synchronized electrical oscillations. 
Likewise, other collectives of bio-oscillators show synchronized oscillations, populations of fireflies being a prominent example \cite{buck1966biology}.
In the case of $\beta$-cells in the pancreas, responsible for the pulsatile release of the hormone insulin, isolated cells do not oscillate or present irregular oscillation patterns~\cite{Perez-1991a}, a fact that points to some collective effect at the origin of the observed coherent oscillations.
Collective effects are known to be fundamental for the concerted working of neurons as well \cite{penn2016network} and also in the case of heart cells the role of non-oscillating cells has been recently revisited~\cite{Maltsev-2022a}.
Building on  related previous work motivated by the applications to $\beta$-cell networks~\cite{Cartwright-2000a,Scialla-2021,Scialla-2022}, the goal of the present article is to revisit some simple models of heterogeneous networks of nonlinear oscillators, such as Fitz{H}ugh-Nagumo  and quartic oscillators, and show their dynamical equivalence to a problem from a very different area of science, the dynamics of a polymer on a one-dimensional substrate.
This equivalence provides an intuitive interpretation of the mechanisms and a simple formulation of the conditions for the appearance of collective oscillations.
\end{quotation}

\section{Introduction}

Understanding the mechanisms underlying synchronization in networks of nonlinear oscillators is an active field of research with numerous applications~\cite{Arenas-2008,Pikovsky-2015a}.
In particular, networks of oscillators have a crucial role in modeling many biological systems.
The human body, for example, contains multiple different networks of cells, including neurons~\cite{Soriano-2023a}, $\beta$-cells in the pancreatic islets of Langerhans~\cite{Peercy-2022a}, and the cardiomyocytes of the heart muscle~\cite{Maltsev-2022a}, all of which present synchronized electrical oscillations.

 The origin of the oscillations in $\beta$-cell networks~\cite{Peercy-2022a} is a long-standing question still without a complete answer.
Various theoretical works have suggested that heterogeneity of $\beta$-cells in the  islets of Langerhans has a key role in producing coherent oscillations.
The complexity of the problem is enhanced by the fact that the consequences of heterogeneity can be very different, ranging from the appearance of synchronization to the inhibition of the coherence of oscillations~\cite{Yamakou-2022}.
The possible effects of heterogeneity on initiating oscillations were pointed out in Ref.~\onlinecite{Smolen-1993a}, where it was shown that diversifying the parameters of a Chay-Keizer model of $\beta$-cell dynamics~\cite{Chay-1983} can lead to synchronized oscillations, whereas the corresponding homogeneous model does not present oscillations.
However, the general meaning of this fact in relation to complex and dynamical systems remained unexplored.

The  heterogeneous $\beta$-cell network model introduced  in Ref.~\onlinecite{Cartwright-2000a} is based on coupled Fitz{H}ugh-Nagumo (FN) oscillators of two different types, characterized by a fixed forcing parameter $f$  that can assume one of two possible values, either  $f_1$ or $f_2$.
In this minimal model, synchronization can appear when some diversity is introduced  in the system, as an emergent process induced by the interactions between these two different types of cells for suitable values of the coupling constant --- whereas the corresponding homogeneous system made up of identical cells would remain in a non-oscillatory state --- as observed experimentally~\cite{Perez-1991a};
synchronization shows a sharp resonance around a specific strength of the coupling between the oscillators.
Another model presenting the appearance of heterogeneity-induced oscillations assigns cells a set of values of the forcing parameter distributed according to a continuum Gaussian distribution~\cite{Tessone-2006a}. This model revealed the existence of an optimal level of diversity, quantified, e.g., by the standard deviation of the Gaussian distribution, at which the network presents the highest coherent response, a phenomenon named diversity-induced resonance (DIR).
Both models present heterogeneity-induced oscillations and a clear resonant behavior of synchronization as some parameter is varied, but using different prescriptions for diversifying the parameters of the oscillators.
Whether the effects induced by a Gaussian distribution~\cite{Tessone-2006a} and a two-value distribution of the bias forces~\cite{Cartwright-2000a} are equivalent or related to each other is still an open problem.

Considering the question from a general dynamical perspective, the first approach to the diversification of the parameters, using a continuum distribution, emphasizes the analogy between DIR and stochastic resonance~\cite{Tessone-2006a,Tessone-2007a}, while the second approach, using a two-value distribution, points to some simple underlying process.
We show that such a process exists and coincides with the resonance effect of a dimer diffusing on a periodic substrate potential: the dimer attains an optimal diffusion rate at a suitable equilibrium rest length (distance between monomers) \cite{Patriarca-2005a, Heinsalu-2008a, Heinsalu-2010a}.
We propose to call such an effect dimer-diffusion resonance (DDR).
As we will discuss, DDR can be generalized straightforwardly to the case of a polymer and in that case it represents the basis of a simple mechanical analog of the appearance of DIR in an oscillator network.   
The DDR mechanism and its extension to polymers is general in nature and is expected to act in a wide category of systems and under different conditions.

\section{Quartic Oscillators}
\label{sec:model}

In this section, we study the synchronization of a heterogeneous network of quartic oscillators.
The results obtained can be directly reapplied also to the case of FN oscillators, considered in Sec.~\ref{sec:FN}.

\subsection{Dynamical equations}

Consider a network of $N$ linearly coupled quartic oscillators evolving according to the equations 
\begin{equation}
    \label{eq:oscillator}
    \dot{x}_i = -V'(x_i) + f(t) +  C \!\! \sum_{j\in\mathcal{N}(i)} (x_j - x_i) + a_i \, ,
\end{equation}
where $i = 1, \dots,N$.
The sum in Eqs.~\eqref{eq:oscillator} represents the interactions between the generic oscillator $i$ and the other oscillators, where a linear coupling of strength $C$ is assumed, extending over the set of oscillators $j \!\in\! \mathcal{N}(i)$ that interact with oscillator $i$.
With $V(x)$ we denote a symmetric double-well potential and $f(t)$ is an external time-periodic forcing. 
In addition, $a_i$ stands for a diversified constant bias force acting on the $i$th oscillator.

Equations \eqref{eq:oscillator} have the same form as the equations that describe $N$ overdamped coupled particles with coordinates $x_i$, moving in an oscillating potential $V(x)-f(t)\,x$, and subject to constant diversified biases $a_i$.
Assuming a suitable rescaling of the space and time coordinates, the potential $V(x)$ in the present case has the symmetric form
\begin{equation}
    \label{eq:quartic}
    V(x) = - \frac{1}{2} x^2 + \frac{1}{4} x^4 \, ,
\end{equation}
with a maximum $V(x = 0) = 0$ and two minima $V(x = \pm 1) = -1/4$.

Differently from excitable oscillators, e.g., Fitz{H}ugh--Nagumo (FN) oscillators, quartic oscillators do not exhibit spontaneous oscillations.
In order to make the oscillatory regime possible, a periodic forcing $f(t)$ is added, with a simple sinusoidal form, an amplitude $b$, and a time period $2\pi/\omega$,
\begin{equation}
    \label{eq:oscillatingforce}   
    f(t) = b \sin(\omega t) \, .
\end{equation}
In the following, we keep the amplitude $b$ constant and small enough that an isolated oscillator cannot oscillate, i.e., the effective potential $[V(x) - x \, f(t)]$ maintains two minima at any time $t$.
We are interested in determining under which conditions  a coupling between the oscillators together with a level of diversification in the set of bias forces $a_i$ induces the appearance of global oscillations in an otherwise silent network. 
To make comparisons consistent with the case of isolated oscillators, for any choice of biases $\{a_i\}$ we assume a zero-mean bias $\langle a \rangle = \sum_i a_i/N \equiv 0$.

 \subsection{Effects of diversification on a fully connected network}
\label{sec:response}

\begin{figure}[tb]
\centering
	\includegraphics[width=\columnwidth]{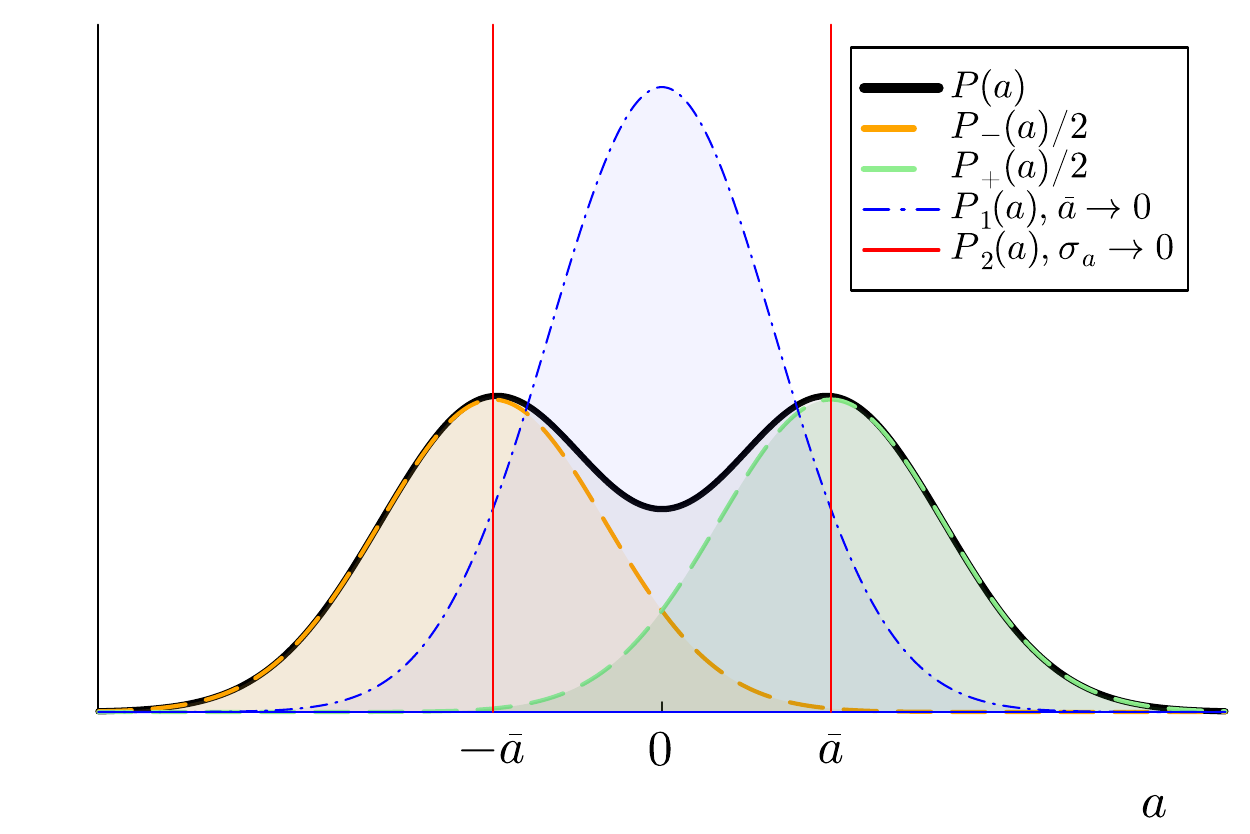}
	\caption{
            Example of bimodal bias distribution function $P(a)$ (black curve) for $\sigma_a/\bar{a} = 2/3$, resulting from the superposition of the partial bias distribution functions $\frac{1}{2} P_\pm(a)$ of the two different types of oscillators (orange and green dashed curves) --- these distributions are defined in Eq.~\eqref{eq:bimodal}. 
            For comparison, we draw also: 
            the limiting Gaussian distribution function $P_1(a)$ (blue dot-dashed curve), given by Eq. \eqref{eq:gaussian}, obtained for $\bar{a} \to 0$ keeping the standard deviation $\sigma_a$ constant; 
            and the two-value $\delta$-distribution function $P_2(a)$ (visualized as two red vertical lines), given by Eq. \eqref{eq:twovalue}, obtained for $\sigma_a \to 0$ keeping $\bar{a}$ constant.  
            }
	\label{fig:bimodal}
\end{figure}

As discussed in the introduction, two possible ways of diversification that can induce an oscillatory regime, are:
(a) the zero-mean Gaussian distribution $P_1(a)$ with standard deviation $\sigma_a$ (used in the study of DIR~\cite{Tessone-2006a}),
\begin{equation}
    \label{eq:gaussian}    
    P_1(a) = \frac{1}{\sqrt{2\pi\sigma_a^2}} \exp\left( - \frac{a^2}{2 \sigma_a^2} \right) \, ;
\end{equation}
(b) the two-point distribution $P_2(a)$ that assigns either a bias $a = -\bar{a}$ or $a = +\bar{a}$ (used in Ref.~\onlinecite{Cartwright-2000a}),
\begin{equation}
    \label{eq:twovalue}    
    P_2(a) = \frac{1}{2} \left[ \delta (a + \bar{a}) + \delta (a - \bar{a}) \right] \, ,
\end{equation}
where $\delta(\cdot)$ is the Dirac delta function.

In the following, we explore the response of an all-to-all connected network of $N$ quartic oscillators to a bimodal distribution of bias forces $a_i$. 
The distribution is assumed to be the superposition of two Gaussian distributions $P_{\pm}(a)$ with mean values $\pm\bar{a}$ and with the same standard deviation $\sigma_a$,
\begin{align}
    \label{eq:bimodal}
    P(a) 
    &= \frac{1}{2} [ P_{-}(a) + P_{+}(a) ]
    \nonumber
    \\
    &\equiv 
    \frac{1}{2 \sqrt{2 \pi \sigma_a^2}} 
        \left\{ 
        \exp\left[-\frac{(a + \bar{a})^2}{2 \sigma_a^2} \right] 
        + 
        \exp\left[-\frac{(a - \bar{a})^2}{2 \sigma_a^2} \right] 
        \right\} \, .
\end{align}
The distribution $P(a)$ is symmetric with respect to $a = 0$ and has the mean value $\langle a \rangle = 0$. 
For generic values $\bar{a}, \sigma_a > 0$, the distribution $P(a)$ represents a hybrid diversification strategy that is intermediate between $P_1(a)$ and $P_2(a)$.
For $\bar{a} \to 0$ the distribution $P(a)$ reduces to the zero-mean normal distribution $P_1(a)$ given by Eq.~\eqref{eq:gaussian}, while for $\sigma_a \to 0$ it becomes the two-point distribution $P_2(a)$ given by Eq.~\eqref{eq:twovalue}. 
An example of the distribution $P(a)$ can be seen in Fig.~\ref{fig:bimodal}.
Applying the distribution $P(a)$ helps to address the question how different diversifications of the constant bias forces $a_i$ ---the two limiting cases given by Eqs.~\eqref{eq:gaussian} and \eqref{eq:twovalue} --- affect the synchronization properties of the oscillator network and provide clues about the origin of the oscillatory regime.
To this aim we explore a fully connected network of $N = 100$ quartic oscillators assuming the following parameter values: rescaled coupling $C = c/N = 0.01$, when choosing $c = 1$;  
tilting amplitude $b = 0.2$ and tilting period $2\pi/\omega = 200.0$;
average bias $\langle a \rangle = \sum_i a_i / N = 0$.

We measure the global response of the system through the quantity $\langle \delta X(t)^2 \rangle$, representing a mean square deviation of the oscillator coordinates averaged in time and over the system oscillators,
\begin{align}
    \label{eq:dX2}
    &\langle \delta X(t)^2 \rangle = \frac{1}{t} \int_0^t ds \left[X(s) - \langle X(s) \rangle \right]^2 \, ,
\end{align}
where
\begin{align}
    \label{eq:dX2_B}
    \langle X(t) \rangle &= \frac{1}{t} \int_0^t ds  X(s) \, ,
    \\
    \label{eq:dX2_C}
    X(t) &= \frac{1}{N} \sum_i x_i(t) \, .
\end{align}
%

\begin{figure*}[tb]
\centering	
    \includegraphics[width=1.1\columnwidth]{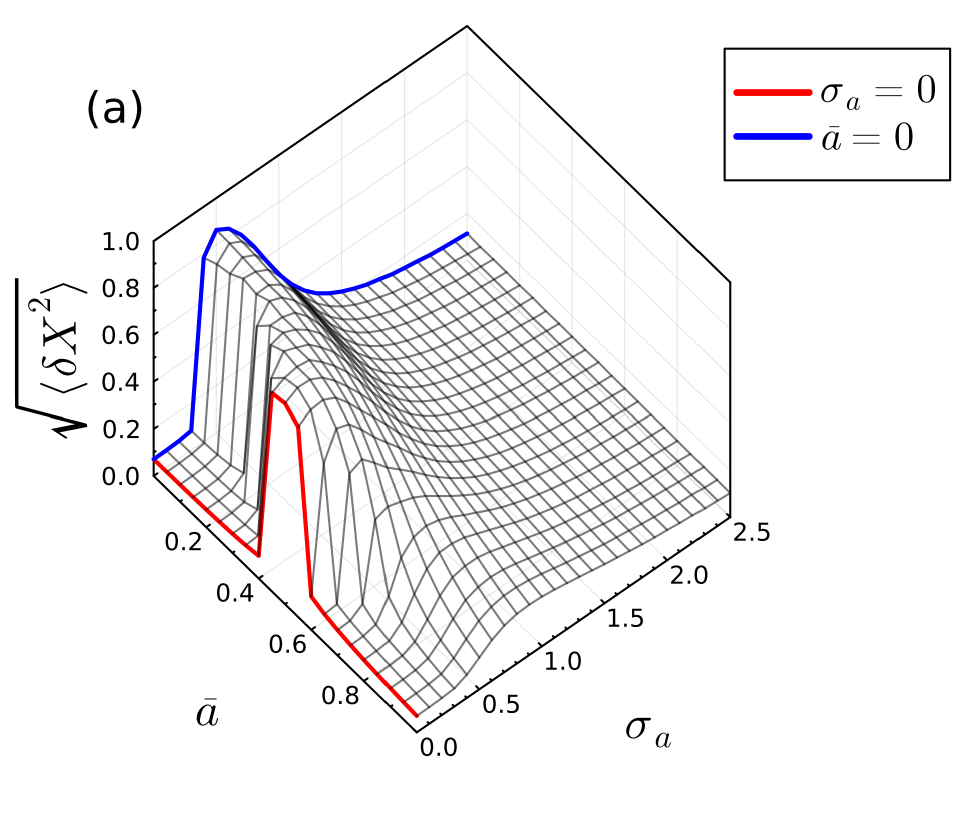}
    \includegraphics[width=0.9\columnwidth]{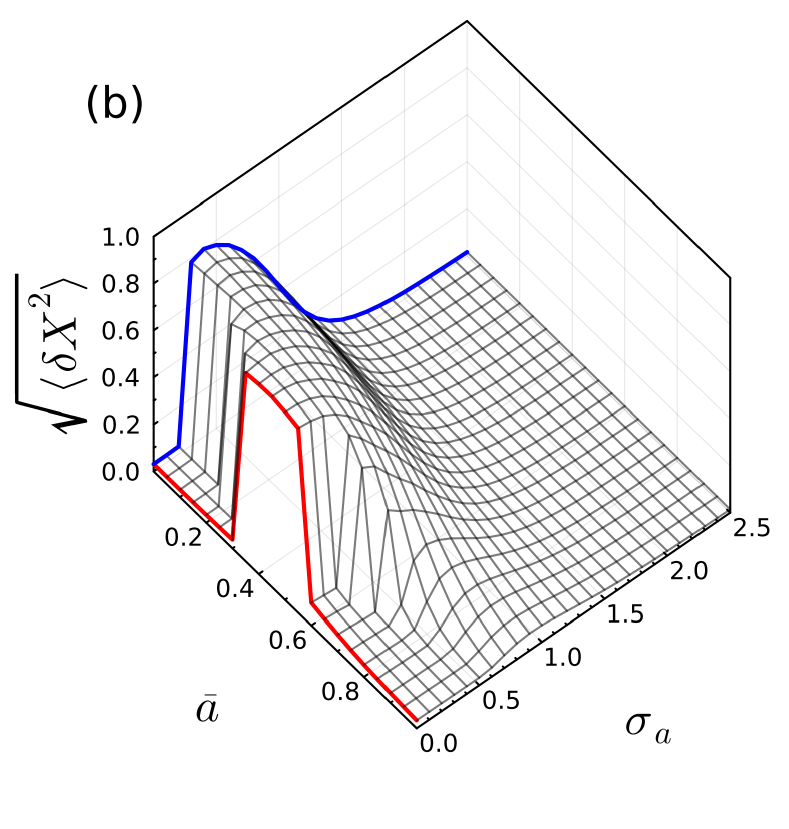}
        \caption{
        Asymptotic global oscillatory activity $\sqrt{\langle\delta X^2\rangle}$ defined in Eqs.~\eqref{eq:dX2}--\eqref{eq:dX2_C} in the $\bar{a}$-$\sigma_a$-plane of a heterogeneous network of \textbf{(a)} quartic oscillators subject to a periodic forcing and \textbf{(b)} FN oscillators.
        The blue curves represent the limit $\bar{a}\to 0$, thus reproducing the results of DIR~\cite{Tessone-2006a}, while the red curves correspond to the limit $\sigma_a \to 0$ for a network with a two-point bias distribution~\cite{Cartwright-2000a}.
        Notice that both these curves have a peak around the value $1/2$.
        The potential $V(x)$ defined in Eq.~\eqref{eq:quartic} is the same for the two types of oscillators, as is the total number of oscillators $N = 100$, the coupling constant $c = 1$, and the final simulation time $t = 2000$. 
        The oscillating force acting on the quartic oscillators, defined in Eq.~\eqref{eq:oscillatingforce}, has period $\tau = 200$ and amplitude $b = 0.2$; the constants regulating the dynamics of the FN slow degrees of freedom $y_i$, Eq.~\eqref{eq:oscillator4}, are $\alpha = 0.02$, $\beta = 0.04$.
        \label{fig:dX2}}
\end{figure*}

The behavior of $\langle \delta X(t)^2 \rangle $ in the $(\bar{a}, \sigma_a)$-plane is shown in Fig.~\ref{fig:dX2}-(a) (we assume $\bar{a} > 0$).
The DIR response of the system is obtained in the limit $\bar{a} \to 0$ and corresponds to the (blue) tick isoline at $\bar{a} = 0$. 
Instead, in the limit $\sigma_a = 0$, the corresponding tick (red) isoline represents the response of the oscillator network to the two-point distribution of the bias $a$.

A relevant feature of the responses depicted in Fig.~\ref{fig:dX2}-(a) is that the isolines obtained in the limiting cases $\bar{a} \to 0$ and $\sigma_{a} \to 0$ are qualitatively similar to each other and both present  a (resonance) peak at the common value $\sigma_a = \bar{a} \approx 1/2$.
Considering that the two curves were obtained using different diversification procedures and are defined using different variables, their similarity suggests a common underlying origin of the respective resonances.
At the same time, there are some important differences, namely the red isoline at $\sigma_a \approx 0$ is sharper, suggesting the existence of a well-defined resonant condition around the value $\bar{a} \approx 0.5$, while the tail of the blue isoline at $\bar{a}=0$ is broader.

\section{Oscillator network as a polymer}
\label{sec:analogy}

\subsection{Dimer-diffusion resonance}
\label{sec:2osc}

In general, the phenomenon of DIR is based on assigning a (Gaussian) distribution of parameters to the single oscillators and thus it is not obvious how to define it for small $N$.
However, it is possible to diversify the constant bias forces even in a small system by assigning to each pair of oscillators $i$ and $j$ opposite biases $a_i = - \bar{a}$ and $a_j = +\bar{a}$. 
This remains valid even in the minimal case of a single pair of oscillators ($N = 2$) described by the equations
\begin{equation}    
    \begin{aligned}
    \label{eq:2oscillators}
    &\dot{x}_1 = - V'(x_1) + f(t) + C \, (x_2 - x_1) - \bar{a} \, , \\
    &\dot{x}_2 = - V'(x_2) + f(t) - C \, (x_2 - x_1) + \bar{a} \, ,
    \end{aligned}
\end{equation}
where the first oscillator is subject to a bias $a = -\bar{a}$ and the second one to a bias $a = +\bar{a}$.
Numerical simulations of this simple two-oscillator system present features analogous to those of the complex network with the two-value bias distribution of Eq.~\eqref{eq:twovalue}, discussed in Sec.~\ref{sec:response}, suggesting that even in the minimal case of the two-particle system described by Eqs.~\eqref{eq:2oscillators} the same mechanism, underlying the global oscillations observed in larger networks, is in action.

It is possible to rewrite Eqs.~\eqref{eq:2oscillators} in the form 
\begin{equation}
\begin{aligned}
    \label{eq:2oscillators-b}
    &\dot{x}_1 = - V'(x_1) + f(t) + C \, (x_2 - x_1 - \ell) \, , \\
    &\dot{x}_2 = - V'(x_2) + f(t) - C \, (x_2 - x_1 - \ell) \, ,
\end{aligned}
\end{equation}
where $\ell = \bar{a}/C$, which shows the equivalence of the two-oscillator problem \eqref{eq:2oscillators} with that of the motion of a harmonic dimer of rest length $\ell$, composed of two monomers with coordinates $x_1$ and $x_2$ linearly coupled with a strength $C$, moving in the potential $V(x) - xf(t)$.
Notice that the dynamic equations \eqref{eq:2oscillators-b} for a dimer corresponding to two coupled quartic oscillator can be derived from an effective potential $W\Qdim(x_1,x_2,t)$,
\begin{equation}
    \label{eq:2oscillators-Weq}
    \dot{x}_1 = -\frac{\partial}{\partial x_1} W\Qdim\!(x_1, x_2, t)\, , \qquad
    \dot{x}_2 = -\frac{\partial}{\partial x_2} W\Qdim\!(x_1, x_2, t)\, , 
\end{equation}
where
\begin{align}
    \label{eq:2oscillators-W}
    W\Qdim\!(x_1, x_2, t) = V(x_1) \!+\! V(x_2) \!-\! (x_1 \!+\! x_2) f(t)
                          + \frac{C}{2} \left(x_2 \!-\! x_1 \!-\! \ell \right)^2_.
\end{align}
Here the last term describes the monomer-monomer interaction within the dimer with equilibrium length $\ell$. 

In Refs.~\onlinecite{Patriarca-2005a,Heinsalu-2008a,Heinsalu-2010a} it was shown that in a spatially periodic potential a dimer exhibits a resonant behavior for an optimal value $\ell^*$ of the rest length $\ell$ close to half  spatial period, at which diffusion and drift under an external force is highest.
We identify this type of resonance, referred to in the introduction as DDR, as the mechanism responsible for the resonant bias observed in Ref.~\onlinecite{Cartwright-2000a} and revealed by the red isoline at $\sigma_a \to 0$ in Fig.~\ref{fig:dX2}-(a).

A simple mechanical explanation of DDR lies in the fact that, for a suitable value $\ell = \ell^*$ of the equilibrium dimer length, the forces acting on the first and second monomer cancel each other and therefore the action of the substrate potential on the dimer is minimized~\cite{Patriarca-2005a,Heinsalu-2008a,Heinsalu-2010a}.
In the particular case of a sinusoidal potential, $V(x) = V_0 \cos(2\pi x/\lambda)$, where $V_0$ and $\lambda$ are the amplitude and spatial period, respectively, the effective amplitude of the periodic potential felt by the center of mass of the dimer is reduced from $V_0$ to $V_0 \cos(\pi(x_2-x_1)/\lambda)$, which implies that when the distance between the monomers is half the spatial period of the potential, $x_2 - x_1 = \ell^* = \lambda/2$, the substrate potential disappears --- see Refs.~\onlinecite{Patriarca-2005a,Heinsalu-2008a,Heinsalu-2010a} for details.

These considerations remain valid when applied to the motion of a dimer in a double-well potential, with the difference that there will be only one resonant rest length $\ell^*$, while in the case of a periodic potential there are infinite resonant lengths $\ell^*_n = \ell^* + n\lambda$ differing by an integer multiple $n$ of the spatial period $\lambda$ \cite{Heinsalu-2008a}.

Notice that the optimal rest length $\ell^* = \bar{a}/c$ depends on the ratio between the bias force and the coupling constant, so that one can equally well study the emergence of an optimal rest length $\ell^*$ fixing the bias $\bar{a}$ and varying the coupling $c$, as done in Ref.~\onlinecite{Cartwright-2000a}, or vice versa fixing $c$ and varying $\bar{a}$, as we do here.

The DDR mechanism also acts in a network of oscillators with $N > 2$, which will be discussed below.

\begin{figure*}[tb]
\centering
    \includegraphics[width=1.6\columnwidth]{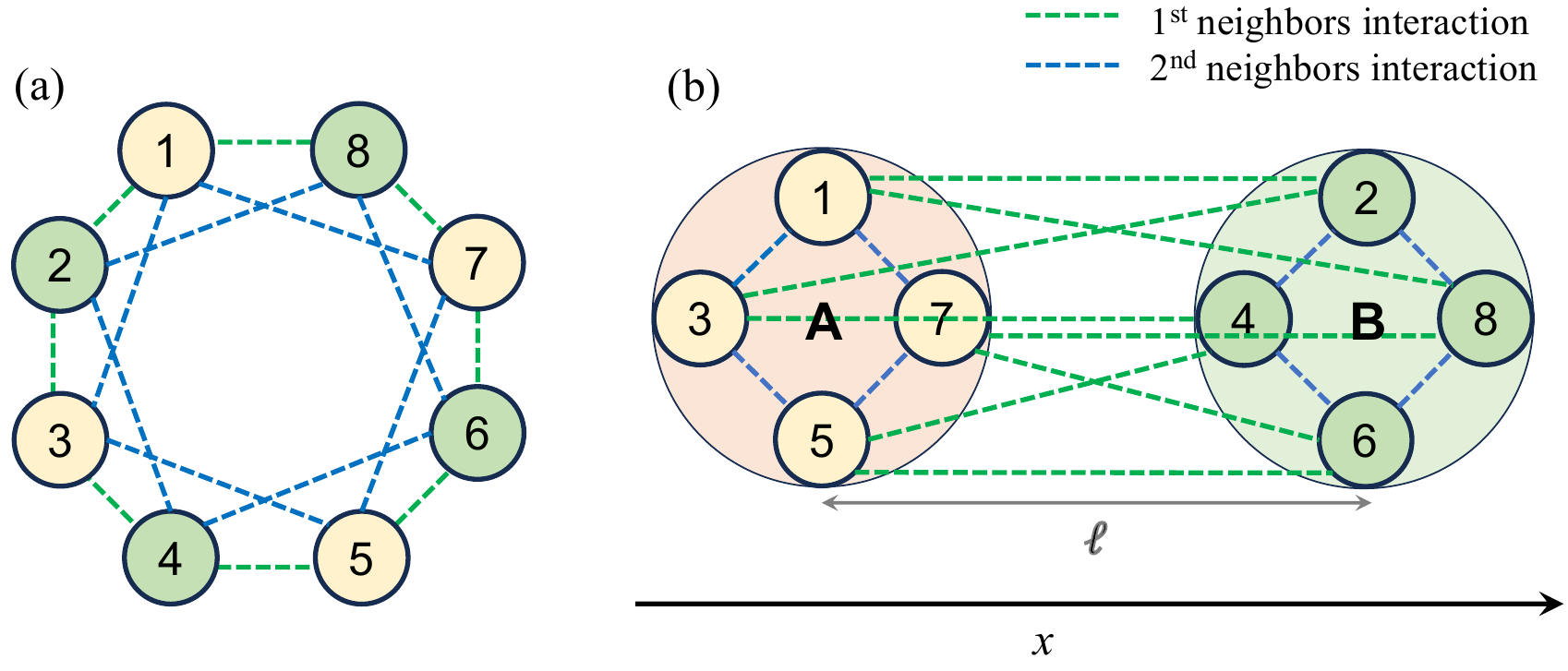}
	\caption{
            \textbf{(a)} Regular network composed of alternating types of oscillators with bias $a = -\bar{a}$ (yellow nodes) and $a = +\bar{a}$ (green nodes). 
            Each node is coupled to two nearest neighbors on both sides, so the degree is $k_0=4$. 
            Blue links represent interactions between two oscillators of the same type, green links between oscillators of different types.
            \textbf{(b)} Polymer mechanical analog in $x$-space. 
            The harmonic forces between particles of the same type tend to induce localized clusters, while harmonic interactions between particles of different type induce the formation of two clusters at a distance $\ell$.
            As a result, the system behaves similarly to two interacting monomers $A$ and $B$ that compose a dimer with equilibrium length $\ell$.  
            }
	\label{fig:clusters}
\end{figure*}

\subsection{Networks with two types of oscillators}

Consider a network composed of an even number $N$ of interacting quartic oscillators, described by Eq.~\eqref{eq:oscillator}, characterized by the two-value bias distribution of Eq.~\eqref{eq:twovalue}, in which a subset $I_{-}$ of $(N/2)$ oscillators is subject to the bias force $a = -\bar{a}$ and the complementary subset $I_{+}$ with the remaining $(N/2)$ oscillators to the opposite bias $a = +\bar{a}$. 
The mean bias in the system is therefore $\langle a \rangle = 0$.
Such a distribution of the bias forces is analogous to that employed in Ref.~\onlinecite{Cartwright-2000a} for the study of a network of FN oscillators;
here it is considered in the framework of quartic oscillators and in the next section in relation to a network of FN oscillators, in order to study when and how it can induce global oscillations.

We divide the system of Eqs.~\eqref{eq:oscillator} into two subsystems corresponding to the oscillator sets $I_\pm$ and indicate the respective coordinates with $x_i^\pm$. 
Correspondingly, also the sum in Eq.~\eqref{eq:oscillator} can be divided into two partial sums.

For the sake of clarity, we consider the case of a regular network where each oscillator has an even number $k_0$ of links equally shared between $(k_0/2)$ oscillators of the set $I_{+}$  and $(k_0/2)$ oscillators of the set $I_{-}$; an example with $k_0 = 4$ is illustrated in Fig.~\ref{fig:clusters}-(a).
We start by introducing a rescaled coupling $c$ in Eq.~\eqref{eq:oscillator},
$$
C = 2c / k_0 \, .
$$ 
Then, for each oscillator $i$, we an can rewrite the corresponding bias $a_i$ as  
$a_i \equiv (2/k_0)\sum_{j = 1}^{k_0/2} a_i$. 
The various terms $(2a_i/k_0)$ can then be absorbed into the linear expressions describing the interaction between oscillator $i$ and the oscillators $j$ of the other type. 
In this way Eqs.~\eqref{eq:2oscillators} can be rewritten as
\begin{equation}
\begin{aligned}
    \label{eq:x+-}
    \dot{x}_i^- &= -V'(x_i^-) + f(t) 
                  +  C \! \sum_{j\in I_{-}} \! (x_j^- - x_i^-) 
                  +  C \! \sum_{j\in I_{+}} \! (x_j^+ - x_i^- - \ell) \, ,  
    \\
    \dot{x}_i^+ &= -V'(x_i^+) + f(t) 
                  +  C \! \sum_{j\in I_{+}} \! (x_j^+ - x_i^+) 
                  -  C \! \sum_{j\in I_{-}} \! (x_i^+ - x_j^- - \ell) \, .  
\end{aligned}
\end{equation}
Here the first sums on the right-hand side represent simple harmonic interactions of oscillator $i$ with oscillators of the same type and the second sums interactions with oscillators of the other type, characterized by an equilibrium distance 
\begin{equation}
    \label{eq:length}
    \ell = \frac{2 \bar{a}}{k_0 C} \equiv \frac{\bar{a}}{c}  \, .
\end{equation}
The dynamical equations can be rewritten with the help of the total potential $W\Qreg(x_1, \dots, x_N, t)$ as
\begin{align}
    \dot{x}_i^- &~= -\frac{\partial}{\partial x_i^-} W\Qreg(x_1, \dots, x_N,t) \, , \label{eq:xW1}\\
    \dot{x}_i^+ &~= -\frac{\partial}{\partial x_i^+} W\Qreg(x_1, \dots, x_N,t) \, ,  \label{eq:xW2}
\end{align}
where
\begin{align}
    \label{eq:WN}
    W\Qreg(x_1, \dots, &\,x_N,t) ~= \sum_{i = 1}^N \! \left[ V(x_i) - x_i \, f(t) \right] 
                          + \frac{C}{2} \sum_{i, j\in I_{-}} \! (x_j - x_i)^2
                          \nonumber
                          \\
    &+ \frac{C}{2} \sum_{i, j\in I_{+}} \! (x_j - x_i)^2
     +  C \! \sum_{i\in I_{-}, j\in I_{+}} \! (x_j - x_i - \ell)^2 \, .
\end{align}
This reformulation of the problem as that of $N$ interacting overdamped particles moving in the total potential $W\Qreg(x_1, \dots, x_N,t)$ suggests a simple mechanical analog of the $N$-oscillator network, namely a polymer moving in a 1D $x$-space, composed of two types of particles, belonging to the sets $I_{-}$ and $I_{+}$.
Pairs of particles of different types interact with each other as monomers of a dimer with an equilibrium length $\ell$ given by Eq.~\eqref{eq:length} and therefore tend to be at a distance $\ell$ from each other (corresponding to the last sum in the total potential in Eq.~\eqref{eq:WN});
instead, pairs of particles of the same type interact through simple harmonic forces and tend to remain as close as possible (second and third sums in Eq.~\eqref{eq:WN}).
As a result, particles of the same type belonging either to $I_{-}$ or $I_{+}$ will form distinct homogeneous localized clusters: a cluster A made up of the particles in $I_{-}$ and another B composed of the particle in $I_{+}$, which will tend to be at a distance $\ell$ from each other; see Fig.~\ref{fig:clusters}-(b). 
Therefore the global response of a $N$-oscillator (regular) network with two-value bias distribution to an external periodic forcing is expected to be similar to that of a single dimer with equilibrium length $\ell$, discussed in Sec.~\ref{sec:2osc}  --- see also Refs.~\onlinecite{Patriarca-2005a,Heinsalu-2008a,Heinsalu-2010a}.
Systems of this type occur naturally, for example in charged dipoles the action of an applied  electric field generates opposite forces on the charged monomers~\cite{Heinsalu-2010a}.   

The dynamical analogy between a network of oscillators and an overdamped polymer can be used to estimate the resonant value $a^*$ of the red isoline at $\sigma_a = 0$ in Fig.~\ref{fig:dX2}-(a).
In a first approximation, we can assimilate the barrier of the bistable potential to one of the barriers of a periodic potential, e.g., a sinusoidal potential. 
Since the potential $V(x) = -x^2/2 + x^4/4$ has two minima $V(x=\pm 1) = -1/4$, their separation $\lambda = 2$ would represent the period of the hypothetical periodic potential, in which it is known that the dimer will exhibit a resonant response when its rest length $\ell$ is equal to half the spatial period $\lambda$~\cite{Patriarca-2005a,Heinsalu-2008a,Heinsalu-2010a}.
Using Eq.~\eqref{eq:length}, we obtain the following approximate resonance condition for a regular network of degree $k_0$,
\begin{equation}
    \label{eq:length2}
    \frac{2 \, \bar{a}}{k_0 \, C} \equiv \frac{\bar{a}}{c} = \frac{\lambda}{2} \, .
\end{equation}
The condition is determined by the ratio $\bar{a}/c$, so that the resonance can be characterized in terms of a resonant bias $\bar{a} = a^*$ or  resonant coupling $c = c^*$.

In the example of the all-to-all connected network of oscillators studied above, with bias distribution given by Eq.~\eqref{eq:twovalue}, discussed in Sec.~\ref{sec:response}, we have $c = 1$ and $N=100$. 
Thus, since a fully connected network is a particular case of a regular network with degree $k_0 = N-1$, from Eq.~\eqref{eq:length2} we obtain a resonant dimer length $a^* =  0.5$, which coincides with the resonant bias value observed in the simulations --- see the red curve in Fig.~\ref{fig:dX2}-(a) obtained in the limit $\sigma_a \to 0$.

\begin{figure}[tb]
\centering
    \includegraphics[width=0.8\columnwidth]{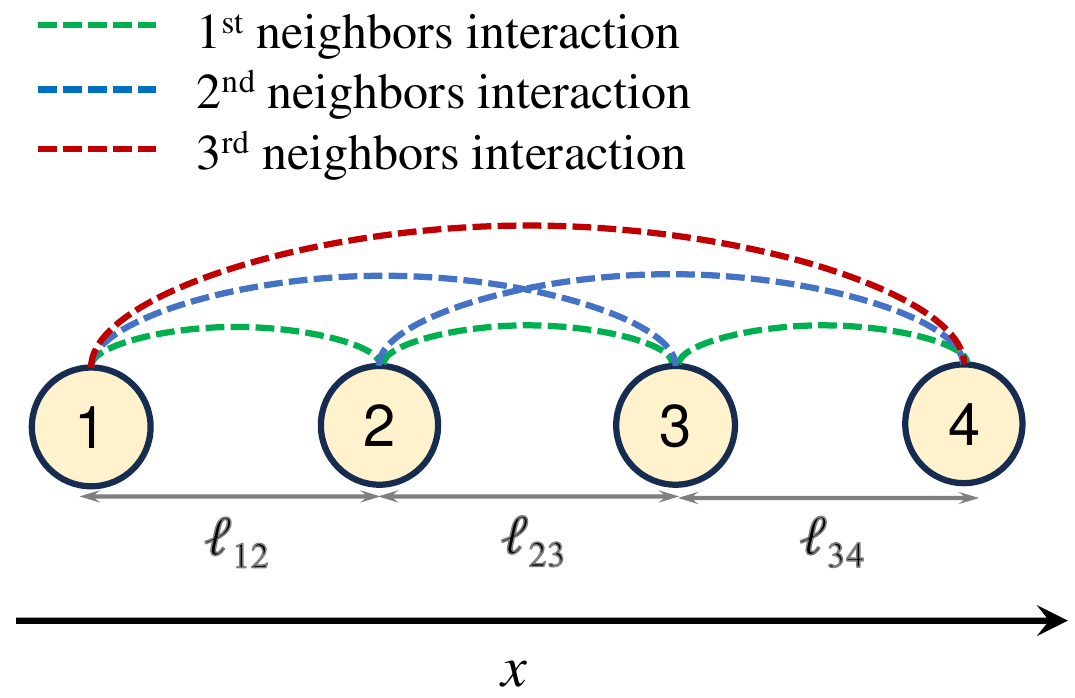}
	\caption{
            Mechanical analog of a small network of four oscillators.
            Each oscillator interacts with all the other oscillators but the equilibrium distances of all the different interactions are consistent with each other according to  Eq.~\eqref{eq:length_i} and produce a robust 1D chain structure.
            }
	\label{fig:linear-polymer}
\end{figure}

\subsection{Network with general bias distribution: diversity-induced resonance}

In this section we show that  DDR takes place in heterogeneous networks with an arbitrary bias distribution and in that case it can be put into correspondence with DIR.

Let us consider the case of an all-to-all connected network, in which the bias values are assigned to the $N$ oscillators according to some continuous distribution $P(a)$, labeling the oscillators in order of increasing bias, i.e., $a_1 < a_2 < \dots < a_N$. 
In order to compare the effects of diversity with respect to a homogeneous network of unbiased oscillators, the only constraint on the distributions is that the mean value is zero,
$\langle a \rangle = N^{-1} \sum_{i=1}^N a_i = 0$.
Then we can rewrite the bias of the $i$th oscillator as
\begin{equation}
    \label{eq:aN}
    a_i = a_i - \langle a \rangle \equiv \frac{1}{N} \sum_{j = 1}^N \left( a_i - a_j \right) \, ,
\end{equation}
and Eqs.~\eqref{eq:oscillator} become
\begin{align}
    \nonumber
    &\dot{x}_i = -V'(x_i) + f(t) 
                + \frac{c}{N} \sum_{j = 1}^N \left( x_j - x_i + \frac{a_i - a_j}{c} \right)
    \\
    \label{eq:oscillator2b}
    &= -V'(x_i) + f(t) 
    - \frac{c}{N} \! \sum_{j=1}^{i-1} \! (x_i -  x_j - \ell_{ij})
    + \frac{c}{N} \! \sum_{j=i+1}^N \! (x_j - x_i - \ell_{j\,i}) \, .
\end{align}
Here we have split the sum into two contributions: 
a sum over oscillators with $j < i$ (therefore with $a_j < a_i$) 
and another sum over oscillators with $j > i$ (with $a_j > a_i$), changing the sign of the first contribution.
In this way all the quantities $\ell_{m\,n}$ in the interaction terms in Eqs.~\eqref{eq:oscillator2b} are positive,
\begin{equation}
    \label{eq:length_i}
    \ell_{m\,n} = \frac{a_m - a_n}{ c } > 0  \quad \text{if} \quad m > n \, ; \quad m, n = 1 \, , \dots, N \, ,
\end{equation}
and can be interpreted as the equilibrium lengths of the corresponding harmonic interaction between the generic $n$th and $m$th oscillators.

The form of the equations above suggests that a polymer represents a mechanical analog of a heterogeneous network, where by ``polymer'' we intend a 1D chain of $N$ mutually interacting monomers with coordinates $\{x_i\}$.
The interactions between monomers are nonlocal, i.e., each monomer $i$ interacts with all the other $(N-1)$ monomers in the system, due to the all-to-all multiple harmonic interactions of the network.
Notice that in this 1D polymer model, monomers will order themselves so that $x_1 < x_2 < \dots < x_{N-1} < x_N$, i.e., in order of increasing bias.
Despite the arbitrariness of the set of bias values $\{a_i\}$, the resulting system is not frustrated, because the various interactions contribute in a consistent way to maintain the same mutual equilibrium distances between monomers and reinforce the global ordered equilibrium structure of the polymer.
This follows directly from the fact that by definition $\ell_{ij} \equiv \ell_{ik} + \ell_{kj}$, for any $i, j, k$, see Eq.~\eqref{eq:length_i}.
For example, the interaction between monomers 1 and 2 has an equilibrium length $\ell_{21}$ and that between monomers 2 and 3 an equilibrium length $\ell_{32}$; but monomer 1 also interacts harmonically with monomer 3, with an equilibrium length given by definition by the right value $\ell_{31} \equiv \ell_{21} + \ell_{32}$ for stabilizing also the 1-2 and 1-3 interactions --- see scheme in Fig.~\ref{fig:linear-polymer}.
This is valid for each of the $N(N-1)/2$ interactions inside the system, since in general the interaction between monomer $i$ and monomer $j$ has an equilibrium length proportional to $|a_i - a_j|$.
Thus, the order of the monomers within the polymer is determined in a unique way by the $N$ values of the bias: from the monomer with the smallest $x$ coordinate, corresponding to the oscillator with the minimum value of the bias, to monomers associated with larger and larger values of bias, until the monomer with the largest coordinate, corresponding to the oscillator with the largest bias.
The larger the number of mutual interactions, the more rigid the structure of the polymer will be, and, eventually, a well-defined configuration of the 1D chain will emerge, with equilibrium distances between two generic monomers $i$ and $j$ given by $\ell_{ij} = |a_i - a_j|/c$. 
The total equilibrium length of the polymer is 
$\ell_\mathrm{tot} = \sum_{i = 1}^{N - 1} \ell_{i, i + 1} = (a_N - a_1)/c$.

\begin{figure}[tb]
\centering
    \includegraphics[width=1.0\columnwidth]{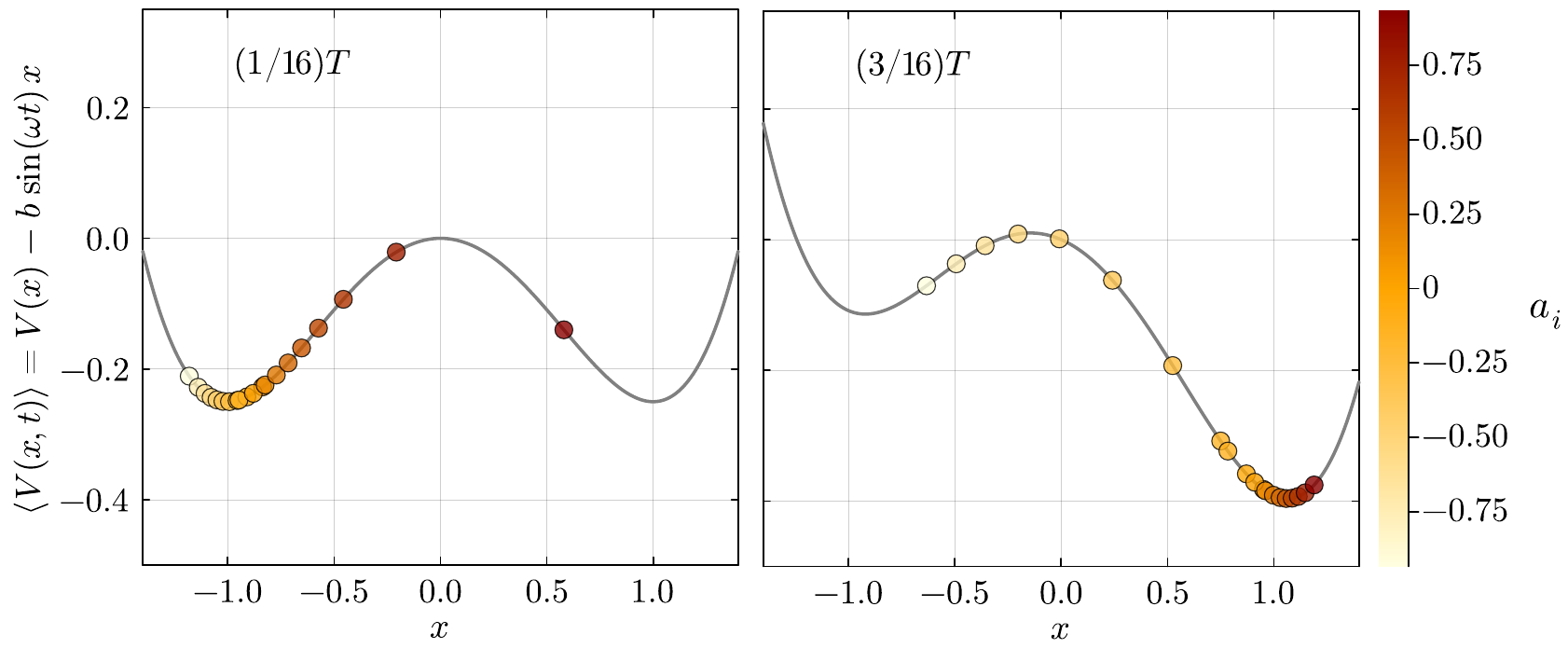}
	\caption{
            Snapshots of the translocation of a polymer composed of $N=20$ monomers, representing a dynamical analog of a network of quartic oscillators subject to a time-periodic forcing defined in Eq.~\eqref{eq:oscillatingforce}. 
            The snapshots are taken at different times within a single time period $T = 2\pi/\omega$. 
            The monomers are color-coded according to the respective values of the external biases $a_i$.
            One can notice that during a polymer translocation (a network oscillation), monomers move in a file maintaining an order based on the bias value, in which the leftmost (rightmost) monomer, with the smallest (largest) $x$-coordinate, has also the smallest (largest) value of the constant external bias.
            The other parameters are as in Fig.~\ref{fig:dX2}.}
	\label{fig:transloc}
\end{figure}

The mechanical analog is apparent by rewriting the equations of motion Eqs. \eqref{eq:oscillator2b} as
\begin{equation}
    \label{eq:oscillator3}
    \dot{x}_i  =  - \frac{\partial}{\partial x_i}    W\Qfull(x_1, \dots, x_N,t) \, ,
\end{equation}
where the effective total potential is simply given by
\begin{align}
  W\Qfull(x_1, ..., x_N,t)  
        \!= \sum_{i} [V(x_i) \!-\! f(t) \, x_i]
        + \frac{C}{2} \sum_{i < j} \left( x_j \!-\! x_i \!-\! \ell_{ji} \right)^2,
        \label{eq:Wgen}
\end{align}
and the sums are extended over all the oscillators.

Numerical simulations show that the collective oscillations of the network correspond to (complete) periodic translocations of the polymer across the potential barrier, from one potential well to the other. 
During the translocation, all the monomers maintain their order in the polymer.
An example of translocation is shown in Fig.~\ref{fig:transloc}.
When the network does not manage to reach a collective oscillatory state, depending on the parameter values, the translocation can be partial, i.e., a part of the polymer remains in the same potential well, or it doesn't take place at all and the polymer remains entirely bound on one side of the potential barrier.
These results confirm the picture that, in DIR scenarios with high levels of diversity, oscillators subject to a bias that is too large in modulus may prevent the whole system from undergoing collective oscillations~\cite{Tessone-2006a}.

Figure~\ref{fig:transloc} also shows that the oscillators perform their oscillations consecutively, one after another, with a finite delay depending on the bias distribution.
In other words, the oscillators, even if oscillating with the same frequency, cannot be in phase.
In this respect, a single collective oscillation resembles the propagation of a pulse along an excitable medium.   

The response of an all-to-all connected oscillator network in the DIR regime is represented by the blue curve in Fig.~\ref{fig:dX2}. 
The DIR peak is located at the same numerical value of the optimal distance, $\sigma_a \approx a^* = 0.5$, because as $\sigma_a$ increases starting from $\sigma_a = 0$, the system will begin to include an appreciable fraction of oscillators characterized by the resonant equilibrium length $a= a^*$ only when the value $\sigma_a \approx a^*$ is reached.
Notice that the DIR response (blue curve) decreases slower than the red curve, a fact that can be expected because, also at $\sigma_a > a^*$, the distribution $P_1(a)$ will include a fraction of values of $a$ around $a=a^*$.
On the contrary, a two-value distribution $P_2(a)$ with a value of $\bar{a}$ appreciably different from $a^*$ will not contain oscillators with equilibrium length around the optimal value.

\subsection{Fitz{H}ugh-Nagumo oscillators}
\label{sec:FN}

All the considerations made above for quartic oscillators also apply to the case of Fitz{H}ugh-Nagumo oscillators~\cite{Fitzhugh-1960a,FitzHugh-1961a,Nagumo-1962a}.
We start from the equations of a single FN oscillator, written in the form
\begin{equation}    
    \begin{aligned}
    \label{eq:FN1}
    &\dot{x} = - V'(x) - y + a \, , \\
    &\dot{y} = \alpha x - \beta y \, ,
    \end{aligned}
\end{equation}
where $V(x)$ is assumed as the same quartic potential Eq. \eqref{eq:quartic} considered above, $\alpha$ and $\beta$ are constants defining the dynamics of the slow degree of freedom $y$, and the constant bias term $a$ appears in the equation for the slow coordinate $x$ instead of the equation of $y$, as in other formulations of the FN dynamics (one can switch between the various formulations through suitable rescaling of the variables and a shift of the $y$ variable).
In this form, it is easier to compare how the DDR mechanism acts within the FN dynamics with respect to the case of quartic oscillators.

Let us consider first the equations of two coupled FN oscillators, linearly coupled in the variables $x_1$ and $x_2$,
\begin{equation}    
    \begin{aligned}
    \label{eq:FN2}
    &\dot{x_1} = - V'(x_1) - y_1 + C(x_2 - x_1) + a_1   \, , \\
    &\dot{x_2} = - V'(x_2) - y_2 - C(x_2 - x_1) + a_2   \, , \\
    &\dot{y_1} = \alpha x_1 - \beta y_1 \, , \\
    &\dot{y_2} = \alpha x_2 - \beta y_2  \, .
    \end{aligned}
\end{equation}
Setting the bias forces of the two oscillators equal in modulus and opposite in sign, i.e., $a_1 = -\bar{a}$ and $a_2 = +\bar{a}$, and introducing the length $\ell = \bar{a}/C$, we can rewrite Eqs.~\eqref{eq:FN2} as
\begin{equation}    
    \begin{aligned}
    \label{eq:FN3}
    &\dot{x_1} = - \frac{\partial}{\partial x_1} W\FNdim(x_1, x_2) - y_1    \, , \\
    &\dot{x_2} =  - \frac{\partial}{\partial x_2} W\FNdim(x_1, x_2) - y_2    \, , \\
    &\dot{y_1} = \alpha x_1 - \beta y_1 \, , \\
    &\dot{y_2} = \alpha x_2 - \beta y_2 \, .
    \end{aligned}
\end{equation}
where the potential $W\FNdim(x_1, x_2)$ is similar to the potential of the quartic oscillators defined in Eq. \eqref{eq:2oscillators-W}, apart from the fact that there is no external oscillating force,
\begin{equation}
    \label{eq:2oscillators-W-FN}
    W\FNdim(x_1, x_2) = V(x_1) + V(x_2) + \frac{C}{2} \left(x_2 - x_1 - \ell \right)^2 \, .
\end{equation}
In fact, also in this case the $x$-sector (the equations for $x_1$ and $x_2$) describes the translational motion in the $x$-space of a dimer of rest length $\ell = \bar{a}/C$, composed of two monomers with coordinates $x_1$ and $x_2$.
In addition, the $y$-sector can be interpreted as describing the internal dynamics of the dimer through the additional coordinates $y_1$ and $y_2$, which produce an alternating tilting force acting on the $x$-degrees of freedom.
In other words, Eqs.~\eqref{eq:FN3} can be interpreted as describing an active dimer moving on the substrate potential $V(x)$.
If the values of the parameters $\alpha$ and $\beta$ are such that the $y$ degree of freedom does not manage to produce a force that pushes the dimer on the other side of the potential barrier, the system will remain in a silent state.
However, the harmonic coupling between the two monomers can drastically change the situation and translocation can take place, with a resonance at a rest length $\ell$ approximately equal to half the distance between the two potential minima.

The above considerations can be generalized to the case of $N$ coupled FN oscillators subject to diversified bias forces $a_i$ ($i = 1, \dots, N$) extracted from an arbitrary distribution $P(a)$ with $\langle a \rangle = 0$, described by the equations
\begin{equation}
\begin{aligned}
    \label{eq:oscillator4}
    \dot{x}_i   &= - \frac{\partial}{\partial x_i} W\FN(x_1, \dots, &\,x_N) - y_i \, ,
    \\
     \dot{y}_i &= \alpha x_i - \beta y_i \, .
\end{aligned}
\end{equation}
For a regular network with degree $k_0$, composed of two types of FN oscillators subject to a bias $a = \pm \bar{a}$, analogous to that depicted in Fig.~\ref{fig:clusters}, the corresponding potential $W\FN=W\FNreg$ is similar to the quartic oscillators potential of the analogous regular network given by Eq.~\eqref{eq:WN}, with the difference that there is no external time-periodic force,
\begin{align}
    \label{eq:W-FN}
    W\FNreg(x_1, \dots, &\,x_N) = \sum_{i = 1}^N V(x_i)
                          + \frac{C}{2} \sum_{i, j\in I_{-}} \! (x_j - x_i)^2
                          \nonumber
                          \\
    &+ \frac{C}{2} \sum_{i, j\in I_{+}} \! (x_j - x_i)^2
     + C \! \sum_{i\in I_{-}, j\in I_{+}} \! (x_j - x_i - \ell)^2 \, .
\end{align}
Also the corresponding resonance condition is unchanged with respect to Eq.~\eqref{eq:length2}.

Finally, in the case of a fully connected network of FN oscillators, with diversified bias forces $a_i$ extracted from a general symmetrical bias distribution $P(a)$, the effective potential $W\FN=W\FNfull(x1,\dots,x_N)$ is similar to the potential defined in Eq.~\eqref{eq:Wgen} of a network of quartic oscillators (apart from the time-periodic force),
\begin{align}
  W\FNfull(x_1, ..., x_N)  
        = \sum_{i} V(x_i)
        + \frac{C}{2} \sum_{i < j} \left( x_j \!-\! x_i \!-\! \ell_{ji} \right)^2.
        \label{eq:WgenFN}
\end{align}
For the latter case, we performed numerical simulations of an all-to-all connected network of $N$ FN oscillators with bias forces $a_i$ diversified according to the bimodal distribution Eq. \eqref{eq:bimodal}.
The response of the system in the $\bar{a}$-$\sigma_a$ parameter plane, measured through its oscillatory activity, is shown in Fig.~\ref{fig:dX2}-(b).
One can notice the close similarity with the response of a system of quartic oscillators, Fig.~\ref{fig:dX2}-(a), which best illustrates the common DDR underlying action.
In fact,  DDR mainly depends on the form of the bistable potential $V(x)$, which has been assumed to be the same in the various oscillator networks considered above, and that is what determines the similarity of their response and, in particular, the similar resonant equilibrium lengths $\ell^*$.

\section{Conclusion}
\label{sec:conclusion}

In this paper we have shown that DDR and DIR are related to each other and have provided evidence that DDR can explain DIR in simple terms.

First, we have shown that a harmonic dimer moving in an external periodic potential is a mechanical equivalent of a system of two coupled bistable oscillators and that the existence of a specific rest length at which the harmonic dimer does not feel the external potential and moves as a free particle is analogous to the resonant oscillatory behavior of the bistable system, observed for a specific value of the modulus of the bias forces acting on the two oscillators.

Then, moving from simpler two-element systems, of more intuitive interpretation, to systems made of $N$ units, we have provided evidence of a connection between a network of bistable or excitable (FN) units and a ``polymer'', i.e., a chain of interacting particles moving on a one-dimensional substrate, which allows one to predict the existence of diversity-induced resonance and to derive analytically the conditions for synchronization. Notably, these predictions are fully consistent with the results of numerical simulations presented here and in previous studies by various authors.

The polymer mechanical analogy allowed us to predict that, also in the case of coupled FN units, one must expect a resonant, oscillatory behavior of the network if the condition Eq.~\eqref{eq:length2} is verified, even when the value of $\bar{a}$ is such that each network element is, individually, in an excitable, i.e., non-oscillatory state. 
This is consistent with the numerical results presented in Ref.~\onlinecite{Cartwright-2000a}. 

The dynamical analogy between nonlinear oscillator networks and 1D polymers offers a novel and general way to study and predict the synchronization properties of many nonlinear systems with a wide range of possible applications, from oscillating biological networks to technological networks.
Also, the same mechanism can be used to explain other similar collective phenomena, such as diversity-enhanced stability~\cite{Ndjomatchoua-2023}.

Although our analysis is based on the assumption that the distribution of bias forces acting on the bistable or excitable units that constitute the network is symmetric  (the  condition usually studied), the same approach can in principle be extended to asymmetric bias distributions, which we will study in the future.

\section*{Acknowledgments}
The authors acknowledge support from the Estonian Research Council through Grant PRG1059.

\section*{Author Declarations}
\subsection*{Conflict of Interest}
The authors have no conflicts to disclose.

\section*{Data Availability Statement}  
The simulation data supporting this study's findings are available from the corresponding author upon reasonable request.







\end{document}